# Natural radioactivity & associated radiological health hazards in soil around Van Eck Power plant, Windhoek, Namibia

Markus Vaefeni Hitila, Sylvanus Ameh Onjefu,

Department of Natural and Applied Sciences, Namibia University of Science and Technology, Windhoek, Namibia

**Abstract**

Primordial radionuclides such as uranium ($^{238}$U), thorium ($^{232}$Th), and potassium ($^{40}$K) and their progenies contained in coal can be a source of concern to the environment in a thermal coal-powered plant. In this study, the average activity concentrations of $^{226}$Ra, $^{232}$Th, and $^{40}$K in the soil around the Van Eck coal-fired power plant in Namibia were determined by the means of the gamma-ray spectrometry technique. The obtained average activity concentrations in the studied soil samples range from 7.74 to 20.04, 8.59 to 31.74, and 108.8 to 484.9 Bq kg$^{-1}$ with an average of 13.33, 17.73, and 269.6 Bq kg$^{-1}$ for $^{226}$Ra, $^{232}$Th, and $^{40}$K, respectively, which were slightly higher than the Windhoek background values. The estimated radiological health hazards were within the prescribed international reference values. The dose rates to which the residents within the 15 km radius are subjected due to combustion activities at the power plant were within the public exposure limits.

## 1. Introduction

The Van Eck Power Plant is the only power plant in Namibia that utilizes coal as a source for energy generation. The power plant has become a great concern for the environment due to the location of the power plant coupled with the nature of the emission of pollutants [1]. Coal contains a trace number of elevated naturally occurring radioactive materials (NORMs) that can be disseminated into the surrounding natural environment during the combustion of coal for power generation [2]. According to [3], the combustion of coal is associated with challenges both from an environmental and human health protection points of view. During the combustion process, these trace elements are mobilized and released through different pathways such as atmospheric emission, and ash deposition of non-combustible elements into the surrounding environment which can lead to an increased concentration of natural radionuclides in the environment with an enrichments factor of 5–10 times the average concentrations of the primordial radionuclides ($^{40}$K, $^{238}$U and $^{232}$Th) [2-4]. The objective of the study was to assess the natural radioactivity concentration in soil within the 15 km radius around Van Eck Coal Power Plant in Namibia and to evaluate the impact on the natural environmental radioactivity.

## 2. Materials and Methods

Soil samples were collected around the Van Eck power plant from 20 sampling points using random sampling. The samples were collected at depth of (0-50 cm) and for each sampling the point, three samples were collected, homogenized to form a composite sample of 1.000 kg, and kept in cleaned and numbered polyethylene bags. Twenty samples were collected at a distance of 5, 10, and 15 km respectively. The samples were dried for seventy-two (72) hours at ambient laboratory temperature to ensure moisture-free samples and then oven-dried at 80 degrees Celsius for 12 hours to attain constant weight. The samples were then thoroughly pulverized, sieved, and homogenized. About 0.5500 kg of the homogenized samples were carefully packed in well-labeled 500 ml Marinelli beakers and hermetically sealed for +30 days to reach secular equilibrium. The radioactivities in the collected samples were measured for 43200 s using a coaxial (62.80 X 64.80 mm) Canberra gamma-ray spectrometer HPGe detector Model No. GC4520 SN 10882 with 45% relative efficiency and resolution of 2.00 KeV (FWHM) at 1.33 MeV peak of $^{60}Co$ and 1.200 keV (FWHM) at 122 keV. The gamma spectrometry system was energy and efficiency calibrated using a multi-nuclide calibration standard with an initial activity of 40kBq homogeneously distributed in silicone matrix, supplied by Eckert & Ziegler Nuclitec GmbH, Germany, SN. AM 5599 and IAEA NORMs reference material RGK-1, RGTh-1, and RGU-1 for $^{40}K$, $^{232}Th$, and $^{238}U$ respectively. The quantification of $^{226}Ra$ activity concentration was done using the gamma energy line of 295.22 keV, 351.93 keV for $^{214}Pb$ and 609.32 keV, 1120.29 keV, and 1764.49 keV for $^{214}Bi$, while 911.21 keV for $^{228}Ac$, 968.97 keV and 238.63 keV for $^{212}Pb$ were used for the assessment of $^{232}Th$ activity concentration. Activity concentration of $^{40}K$ was evaluated using the single 1460 keV gamma-line of potassium.

## 1. Results and discussion

### 1.1. Activity concentrations of $^{226}Ra$, $^{232}Th$, and $^{40}K$

The average activity concentrations of the studied radionuclides are presented in Table 1. The average activity concentrations for $^{226}Ra$, $^{232}Th$, and $^{40}K$ were calculated and found to be 13.58 ± 3.33, 17.73 ± 5.88, and 269.6 ± 40.72 Bq/kg respectively. Although the obtained activities were above the average natural environment background for Windhoek, they fall within the range of world average values of 35, 30, and 400 Bq/kg for $^{226}Ra$, $^{232}Th$, and $^{40}K$, respectively. The activity concentrations of $^{226}Ra$ and $^{40}K$ showed a relatively flat distribution while $^{232}Th$ had a steeper than a normal distribution. However, a positively skewed concentration was obtained which implies that the distribution of radionuclides within the studied soil samples was asymmetric around the mean. The high sample variance observed for $^{40}K$ is because of the wide range observed in its activity concentrations in the 20 soil samples collected. The even distribution of $^{226}Ra$, $^{232}Th$, and $^{40}K$ is attested by the frequency distributions histograms shown in Figure 1.

**Table 1:** Summary statistics for activity concentrations of $^{226}Ra$, $^{232}Th$, and $^{40}K$, in soil samples within the 15 km of Van Eck power plant.

| **Descriptive Statistics** | **$^{226}Ra$ [Bq/kg]** | **$^{232}Th$ [Bq/kg]** | **$^{40}K$ [Bq/kg]** |
|---|---|---|---|
| **Minimum** | 7.24 | 8.59 | 108.83 |
| **Maximum** | 20.04 | 31.74 | 484.98 |
| **Mean ± SD** | 13.58 ± 3.33 | 17.73 ± 5.88 | 269.6 ± 40.72 |
| **skewness** | 0.24 | 0.74 | 0.18 |
| **Kurtosis** | -0.47 | 0.05 | -1.11 |
| **Sample Variance** | 10.75 | 38.10 | 13856.95 |
| **Windhoek, Namibia [5]** | **8.54** | **15.80** | **240.00** |
| **World Average Value** | **35** | **30** | **400** |

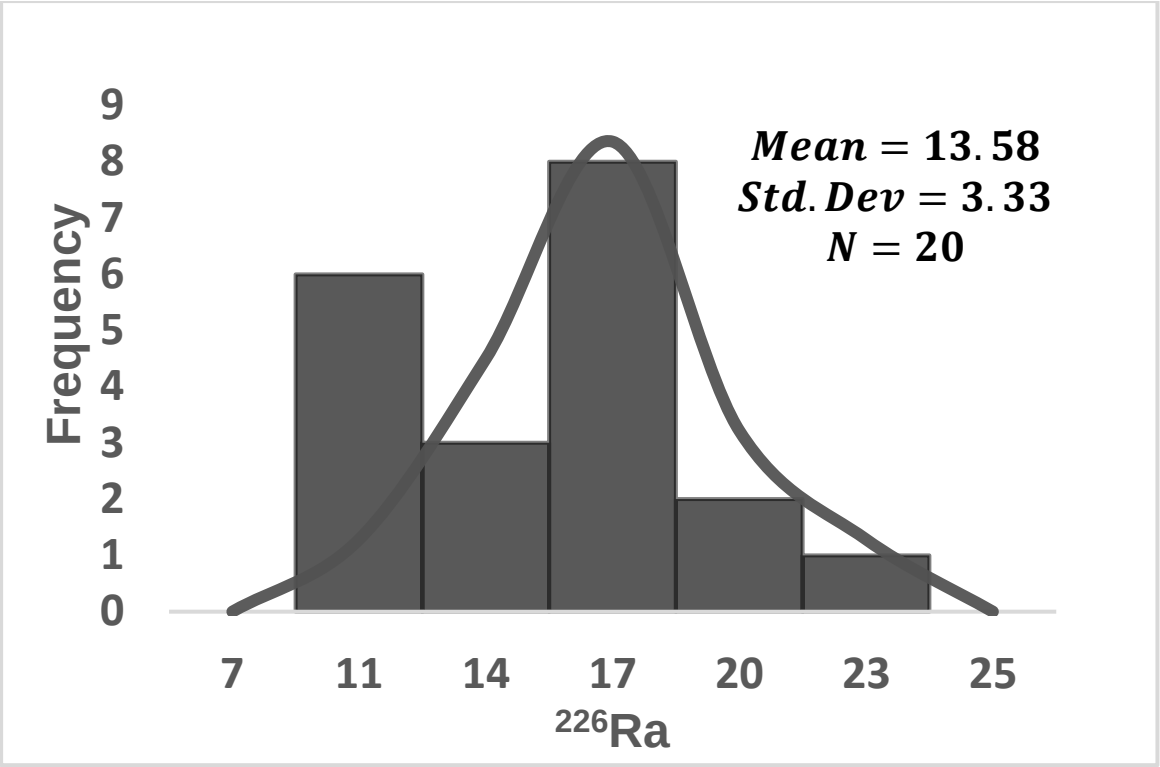

Fig. 1a

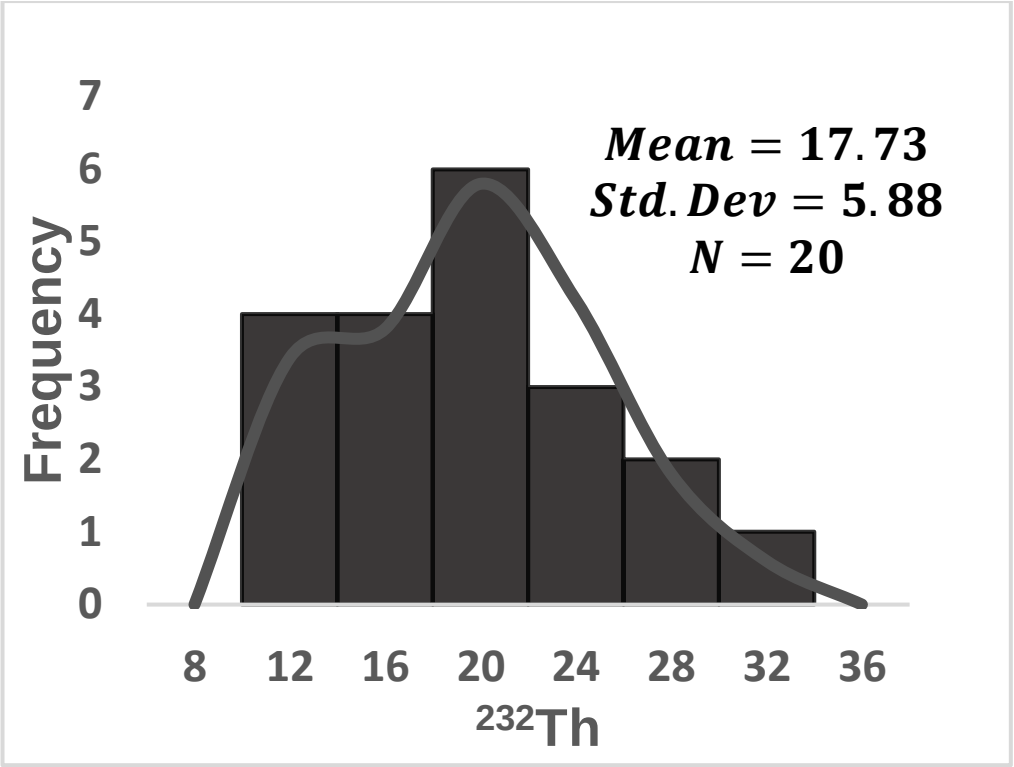

Fig. 1b

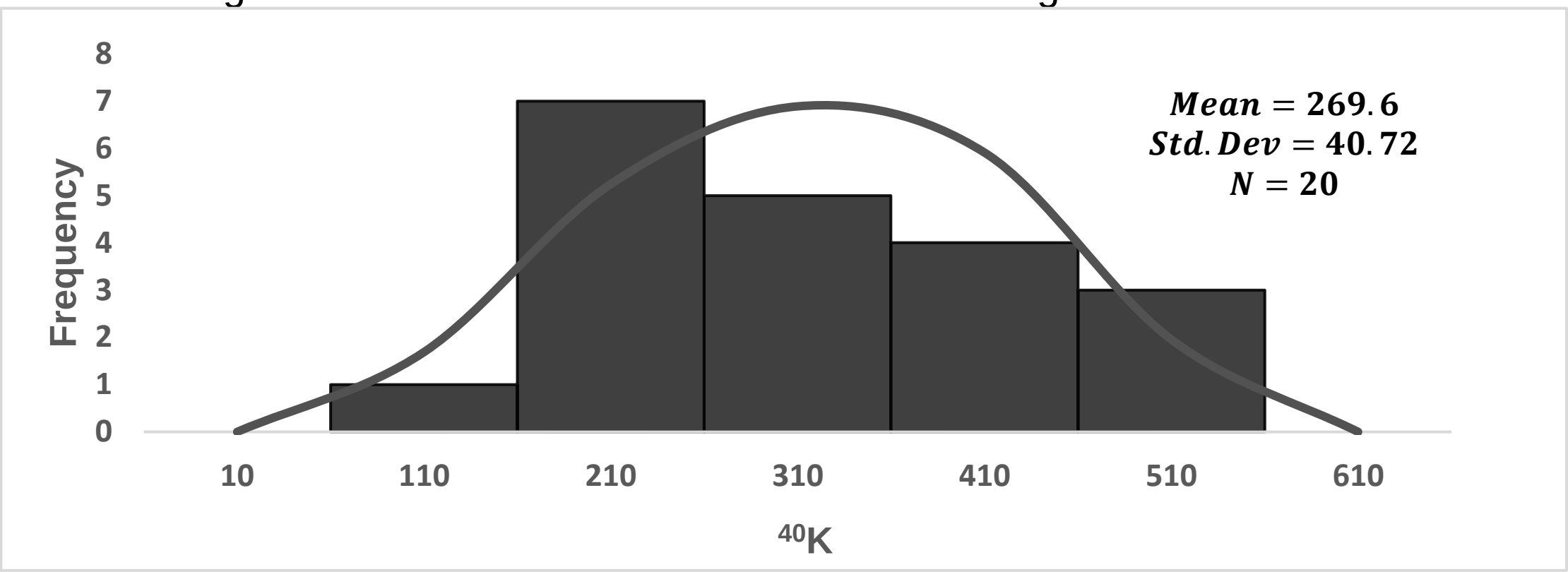

Fig. 1c

Figure 1: Frequency distribution histogram of (a) $^{226}$Ra, (b) $^{232}$Th, and (c) $^{40}$K in soil sample around Van Eck Coal Power Plant.

## 2.1. Radiological hazards assessments

The radiological health hazards such as radium equivalent activity ($Ra_{eq}$), the external hazard index ($H_{ex}$), the internal hazard index ($H_{in}$), annual effective dose equivalent (AEDE), absorbed dose rate $D_{R,}$ representative gamma index ($I_{\gamma}$), and excess lifetime cancer risk (ELCR) were estimated and presented as per Table 2.

The $Ra_{eq}$ of the sample containing different levels of $^{226}$Ra, $^{232}$Th, and $^{40}$K nuclides was evaluated using equation 1, and the permissible limit of $Ra_{eq}$ required to keep the absorbed dose < 1.5 mGy/h, must be < 370 Bq/kg [3].

$$Ra_{eq}(Bqkg^{-1}) = A_{Ra} + 1.43A_{Th} + 0.077A_K \quad (1)$$

The quantification of the incurred radiation hazard due to external exposure to gamma-rays and the administered radiation hazards to the sensitive organs from radon and its short-lived radionuclides was assessed as follows [6]

$$H_{ex} = \frac{A_{Ra}}{370} + \frac{A_{Th}}{259} + \frac{A_K}{4810} \leq 1 \quad (2)$$

$$H_{in} = \frac{A_{Ra}}{185} + \frac{A_{Th}}{259} + \frac{A_K}{4810} \leq 1 \quad (3)$$

The gamma radiation hazard due to the respective concentration of the investigated natural radionuclides was assessed by the representative gamma index. The index serves as a screening parameter for the material of possible radiation health challenges [2].

$$I_{\gamma} = \frac{A_{Ra}}{150} + \frac{A_{Th}}{100} + \frac{A_K}{1500} \leq 1 \quad (4)$$

The outdoor and indoor AEDE in mSv/y from the radioactivity content of the sample was calculated by applying two conversion coefficients provided by [6] as follows:

$$AEDE(mSvy^{-1}) = D_R(nGyhr^{-1}) \times 8760\frac{h}{y} \times \frac{10^{-6}}{10^{-9}}\frac{mGy}{Gy} \times 0.7 \times 0.2\frac{Sv}{Gy} \quad (5)$$

For the uniform distribution of radionuclides, the gamma absorbed dose rate 1 m above the air was calculated as per [6] guidelines.

$$D_R(nGyh^{-1}) = 0.462A_{Ra} + 0.604A_{Th} + 0.0417A_K \quad (6)$$

The excess lifetime cancer risk (ELCR) for radiation exposure from soils was estimated as follows [7].

$$ELCR = AEDE \times DL \times RF \quad (7)$$

where $A_{Ra}$, $A_{Th}$, $A_K$, AEDE, DL, and RF are the activity concentrations of $^{226}Ra$, $^{232}Th$, and $^{40}K$, annual effective dose equivalent, duration of life (70 years), and risk factor (0.05 $Sv^{-1}$ for stochastic effects for the public were used) respectively.

The estimated values for studied radiological health hazards ranged from 28.57 to 102.77 Bq/kg for $Ra_{eq}$, 13.78 to 50.52 nGy/h for $D_R$, 0.02 to 0.06 mSv/y for AEDE, 0.08 to 0.28 for $H_{ex}$, 0.10 to 0.33 for $H_{in}$, 0.14 to 0.45 for $I_\gamma$ and 0.06 x $10^{-3}$ to 0.22 x $10^{-3}$ for ELCR. The health hazards indicators were all found to be within the acceptable international safety limits.

The correlation coefficients of $^{226}Ra$, $^{232}Th$, and $^{40}K$ for the 95 % confidence indicated a strong positive linear correlation between the measure activities and the absorbed dose rate as shown in figures 2-3, signifying that the impact of radiation on the environment around the power plant is evenly distributed among the studied radionuclides. Also, figures 4 and 5 indicated a strong positive correlation between the radionuclides of $^{40}K$ and $^{226}Ra$ and between $^{40}K$ and $^{232}Th$ respectively which indicates similar source and behaviour in the environment. These can be attributed to the geology of Windhoek, which is characterized by quartzite, ubordinate calcareous schist, impure marble, and amphibole schist of the Damara sequence subgroups. This formation is characterized by radiogenic minerals.

**Table 2**: Summary statistics of dose and radiation hazard indexes of soil around and within the vicinity of the power plant.

| Descriptive statistics | $Ra_{eq}$ [Bq/kg] | $D_R$ [nGy/h] | AEDE [mSv/y] | $H_{ex}$ | $H_{in}$ | $I_\gamma$ | ELCR [$\times 10^{-3}$] |
|---|---|---|---|---|---|---|---|
| Min | 28.57 | 13.78 | 0.02 | 0.08 | 0.10 | 0.14 | 0.06 |
| Max | 102.77 | 50.52 | 0.06 | 0.28 | 0.33 | 0.45 | 0.22 |
| Mean | 59.50 | 29.15 | 0.04 | 0.16 | 0.20 | 0.26 | 0.13 |
| World Permissible limits | < 370 | 59.00 | 0.07 | ≤ 1 | ≤ 1 | ≤ 1 | 0.29 [0.05] |

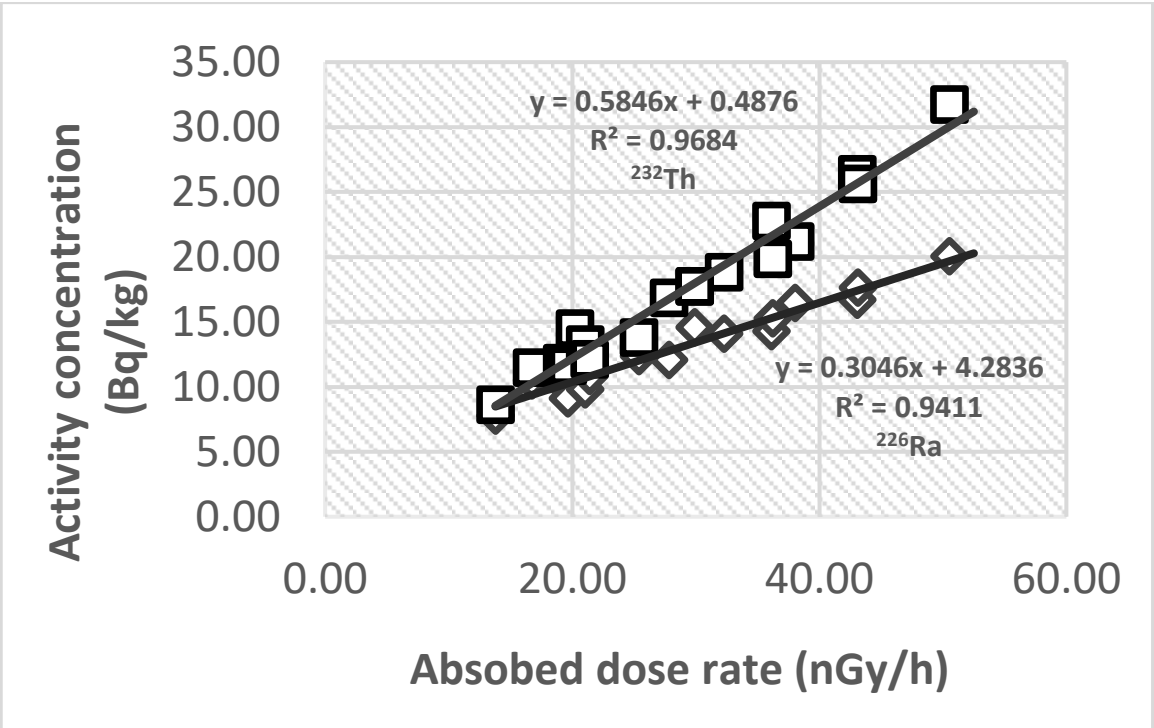

Figure 2: Correlation analysis between activity of $^{226}Ra$, $^{232}Th$, and absorbed dose rate.

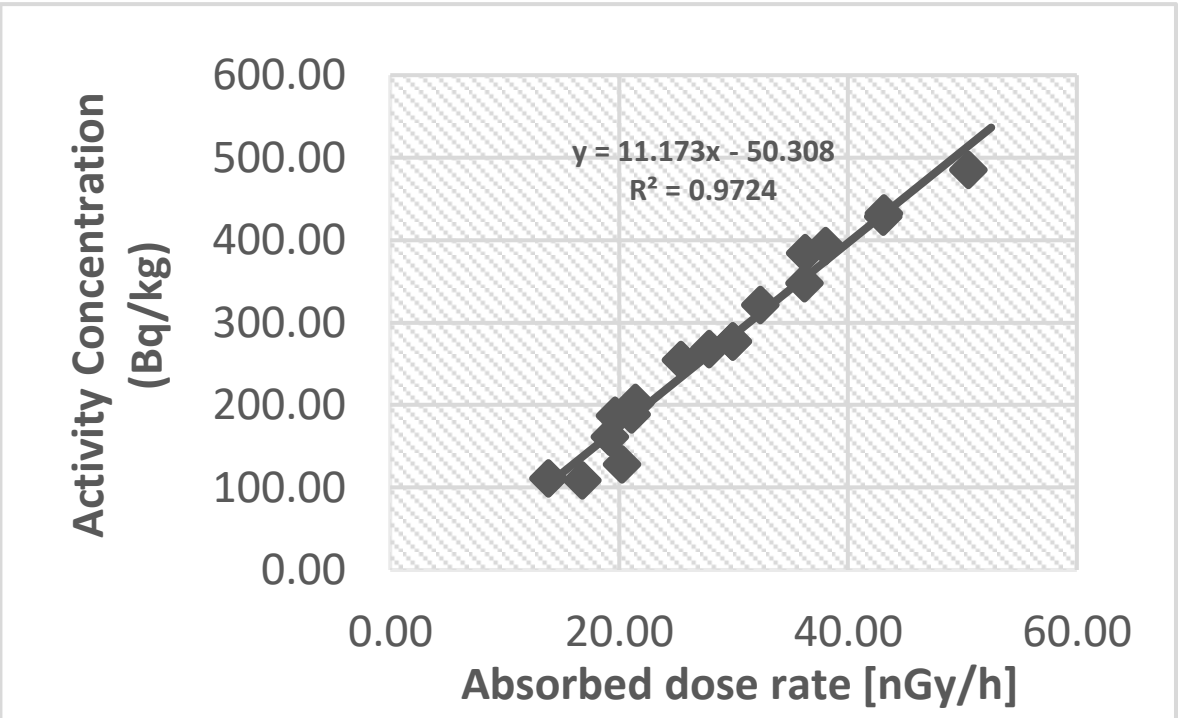

Figure 3: Correlation between the activity of $^{40}K$ and absorbed dose rate.

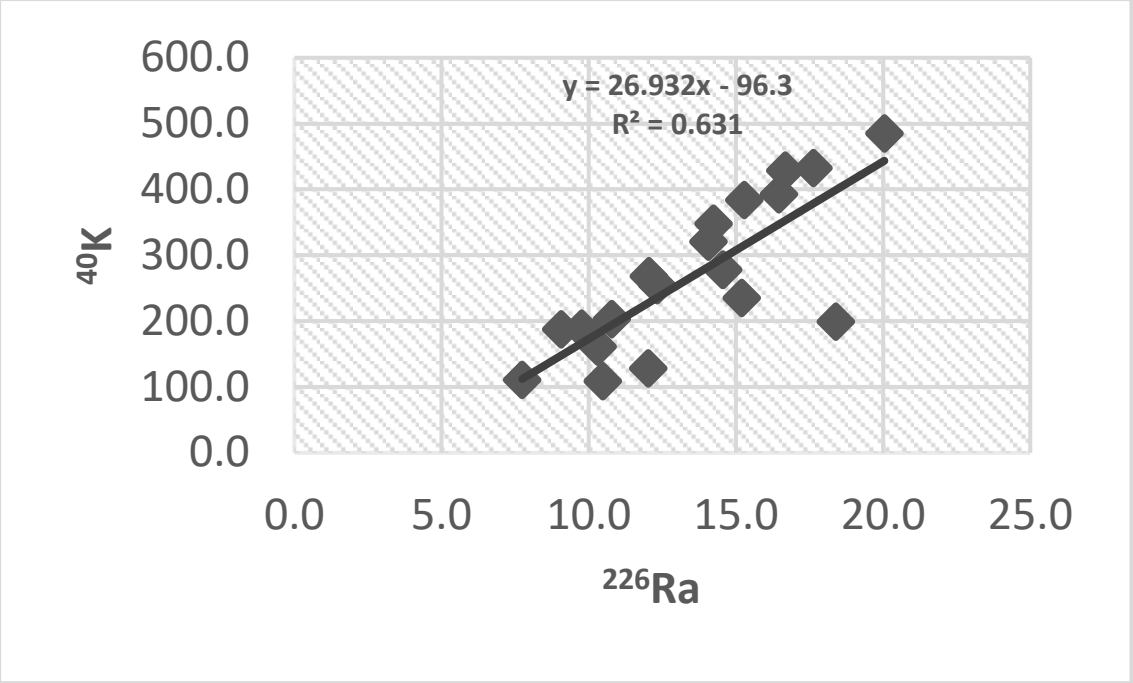

Figure 4: Correlation of $^{40}K$ and $^{226}Ra$

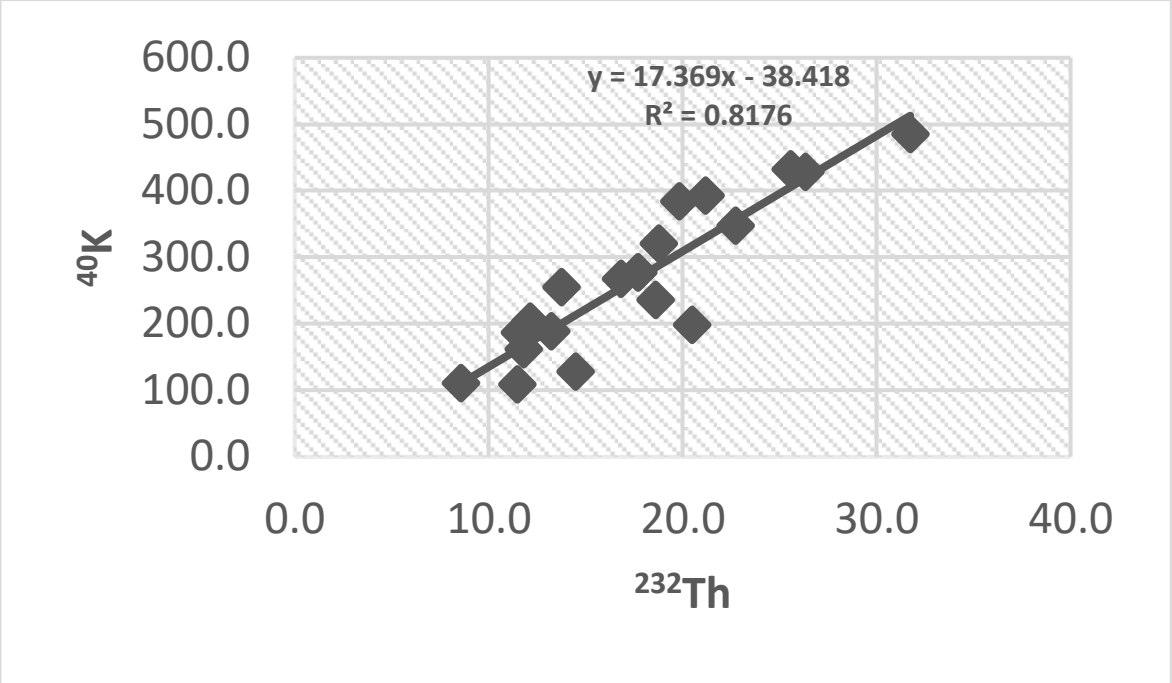

Figure 5: Correlation of $^{40}K$ and $^{232}Th$

## 4. Conclusion

The environmental radioactivity impact of the Van Eck Power Plant has been assessed in this study. The obtained average activity concentrations of $^{226}Ra$, $^{232}Th$, and $^{40}K$ in the investigated samples were found to be within the world average values. The estimated radiological health hazards from the measured activity concentration were below the international acceptable limit of safety. The statistical analysis showed that the distribution of gamma exposure to the population within and around the power plant is evenly distributed among the assessed radionuclides. The study shows that the combustion of coal has a noticeable influence on the natural environmental radioactivity by enhancing the natural radiation in the surrounding environment. Hence, the management of residual waste and monitoring of emissions should be enhanced to mitigate any adverse effects on the surrounding environment.

**Acknowledgment**

The authors would like to extend their gratitude to Van Eck power Plant management and staff for their cooperation in gathering samples and sharing the necessary technical data.